\documentclass[pre,twocolumn,aps,showpacs,superscriptaddress]{revtex4}
\pdfoutput=1
\usepackage{graphics}
\usepackage{epsfig}
\usepackage{amsmath}
\usepackage{amssymb}

\begin{document}
\title{Force balance in canonical ensembles of static granular packings }

\author{Brian P Tighe}
\affiliation{Instituut-Lorentz, Universiteit Leiden, Postbus 9506, 2300 RA Leiden, The Netherlands}
\author{Thijs JH Vlugt}
\affiliation{Delft University of Technology, Process \& Energy Laboratory, Leeghwaterstraat 44, 2628 CA Delft, The Netherlands}

\date{\today}

\begin{abstract}
We investigate the role of local force balance in the transition from a microcanonical ensemble of static granular packings, characterized by an invariant stress, to a canonical ensemble. Packings in two dimensions admit a reciprocal tiling, and a collective effect of force balance is that the area of this tiling is also invariant in a microcanonical ensemble. We present analytical relations between stress, tiling area and tiling area fluctuations, and show that a canonical ensemble can be characterized by an intensive thermodynamic parameter conjugate to one or the other. We test the equivalence of different ensembles through the first canonical simulations of the force network ensemble, a model system.
\end{abstract}
\pacs{45.70.Cc, 05.40.–a, 46.65.+g}
\vspace{1cm}

\maketitle

Dense packings of grains are athermal: having achieved a mechanically stable state, they remain there unless externally driven. The set of final states of a particular numerical or experimental preparation protocol defines a nonequilibrium ensemble of static packings. A fundamental and open question, ultimately related to the packing and mechanical properties of granular media, is the proper framework for the statistical description of such an ensemble. This work addresses one important aspect of the problem, namely the role of local force balance and how it influences the transition from a microcanonical ensemble, here defined as an ensemble at fixed stress, to a canonical ensemble.

Energy is a natural quantity to characterize equilibrium states because it is an invariant of the dynamics. This distinction is lost in an ensemble of static packings, and one must search for other convenient quantities to replace it. A number of prior works have proposed characterizing a statistically homogeneous packing by its average stress $\hat \sigma$ \cite{evesque99,kruyt02,ngan03,bagi03,goddard04,metzger05,henkes07,metzger08,edwards08}.  This is a natural choice in the sense that the ``extensive stress'' ${\hat \sigma}V$ 
\footnote{
We abuse units for linguistic convenience. Here and throughout, ``extensive stress'' and ``extensive pressure'' have units ${\rm force} \times {\rm length}$.
} on a body of volume $V$ in mechanical equilibrium under an external load is determined by the forces on its boundary, regardless of the configuration of grains in the bulk \cite{landau}. A microcanonical ensemble at fixed extensive stress may then be defined as the set of all arrangements of $N$ grains consistent with a particular boundary loading. Throughout this work we restrict ourselves to frictionless packings of disks in two dimensions and to isotropic stress states, so that the extensive pressure $\cal P$ = $\frac{1}{2}({\rm Tr}\,{\hat \sigma})V$ suffices to characterize the stress. The extensive pressure is additive: ${\cal P} = \sum_i p_i$. In a disk packing the ``pressure'' on a grain is $p_i = \sum_j {\vec f}_{ij} \cdot {\vec r}_{ij}$, where ${\vec f}_{ij}$, the force that grain $j$ applies to grain $i$, is nonzero only if the grains are in contact and ${\vec r}_{ij}$ is the vector from the center of $j$ to $i$.

Postulating entropy maximization, a Boltzmann-like factor $\exp{(-\alpha {\cal P})}$ follows for the canonical ensemble, provided $\cal P$ is the only relevant invariant of the microcanonical ensemble. The quantity $\alpha$ is an intensive thermodynamic parameter conjugate to $\cal P$. Edwards has recently suggested the name {\em angoricity} for $\alpha^{-1}$ \cite{edwards08}. It is distinct from the more well-known compactivity \cite{edwards89}, which is conjugate to the packing volume $V$. 

To date, the main application of stress-based ensembles of static packings has been the prediction of the statistics of local measures of stress, such as the force $f$ at a contact or the pressure $p$ on a grain, which provide a fundamental characterization of the stresses in a packing. The forces on one grain are coupled to the forces on other grains via Newton's third law. In a statistical treatment, local force balance enters through $\delta$-functions in the partition function of the system \cite{snoeijer04b}, which must then be integrated over; in analytical calculations this is tedious or impossible for systems larger than a few grains \cite{snoeijer04b,tighe05}. All published calculations resort to some degree of approximation to treat local force balance, and most neglect spatial correlations completely. At the latter level of approximation, the tail of a local stress probability distribution function reflects the form of the Boltzmann factor, and hence these approaches predict local stress probability distribution functions that decay exponentially for large stresses \cite{evesque99,kruyt02,bagi03,goddard04,edwards08,coppersmith96}. This prediction is analogous to the Maxwell-Boltzmann distribution in an ideal gas, which is exponential in the particle energy. We will refer to calculations that neglect spatial correlations as ``ideal gas-like'', though this is not meant to invoke a gaseous state.

In recent work \cite{tighe08b}, we pointed out that it is possible to improve on ideal gas-like calculations in 2D packings by making use of a dual structure known as the Maxwell-Cremona diagram or reciprocal tiling \cite{maxwell1864} . In Section \ref{sec:tiling}, we construct the reciprocal tiling and explain how it influences local stress statistics. Crucially, the tiling exists as a necessary consequence of local force balance. The most important feature of these tilings is their area $\cal A$, which we shall see is an extensive invariant much like $\cal P$. As the constraint of mechanical equilibrium is one of the principal differences between ensembles of granular packings and other ensembles, a key question is how the tiling area $\cal A$ can be incorporated in an ensemble treatment of static granular packings beyond the ideal gas approximation. The remainder of this work seeks to answer this question. 

\section{Reciprocal tilings}
\label{sec:tiling}
\begin{figure}[tbp] 
\centering
\includegraphics[clip,width=0.9\linewidth]{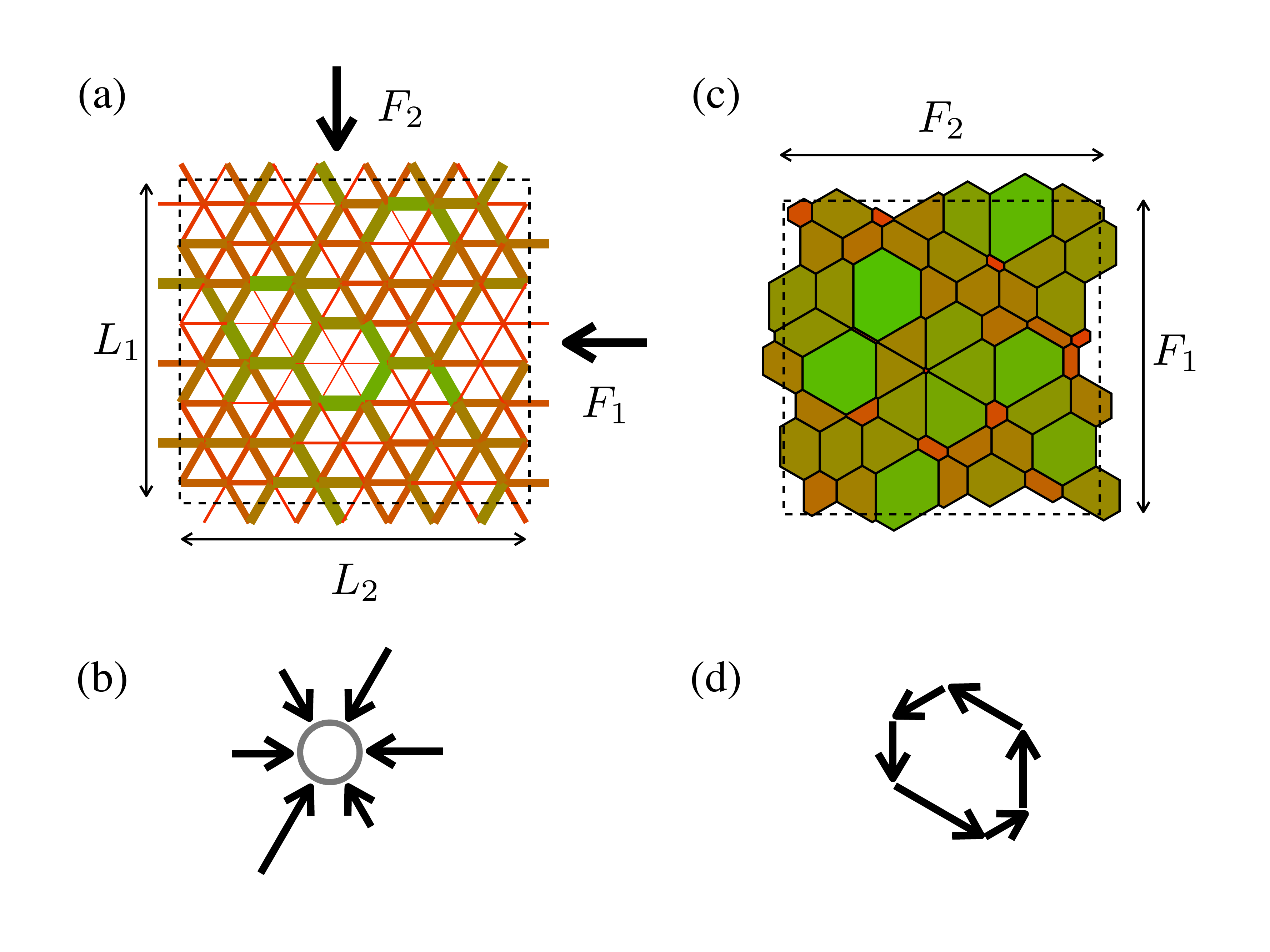}
\caption{ (a) A network of balanced contact forces on the  frictionless periodic triangular lattice. Nodes correspond to grains and edges to contacts. Color and line thickness are mapped to contact force magnitude. The contact network has dimensions $L_1 \times L_2$. The total force on a surface parallel to the boundary of the unit cell with dimension $L_1$ ($L_2$) is ${\vec F}_1$ (${\vec F}_2$). (b) A single-grain state in the triangular lattice. Arrows represent vector contact forces acting on the grain. (c) The reciprocal tiling, or Maxwell-Cremona diagram, corresponding to the force network in (a). Each tile is constructed from the contact forces acting on a grain. The unit cell of the tiling has dimensions $F_1 \times F_2$.	 (d) Construction of a tile for the grain in (b). Vector forces, rotated by $\pi/2$, are graphically summed around the grain; the tile closes because the grain is in static force balance. The tiles tessellate space due to Newton's third law. Rotating forces by $\pi/2$ is not essential, but doing so ensures that, e.g., if grain $j$ is to the right of grain $i$ in the packing, tile $j$ will be to the right of tile $i$ in the tiling.}
\label{fig:tiling}
\end{figure}
The Maxwell-Cremona diagram or reciprocal tiling is a dual structure constructed from the forces in a two-dimensional packing; it is a geometric representation of the stress state (see Fig.~\ref{fig:tiling}a,c). Each grain maps to an individual tile in the tiling. The boundaries of this tile are constructed by graphically summing the contact forces on the grain, moving from contact to contact around the grain in a right hand fashion (see Fig.~\ref{fig:tiling}b,d). A tile's boundary closes because the vector sum of the forces on the grain is zero, i.e.~because the grain is in force balance. Moreover, tiles tessellate space due to Newton's third law, which guarantees tiles in contact have facets with like length and orientation. Note that lengths in the tiling correspond to forces in the packing, so the tiling occupies a different space than the packing. In particular, the area $a_i$ of the tile corresponding to grain $i$ has units of $({\rm force})^2$. Modulo a global rotation, the vertex coordinates in a Maxwell-Cremona diagram are equivalent to the ``void forces'' of Satake \cite{satake93} or ``loop forces'' of Ball and Blumenfeld \cite{ball02}. 

As noted above, the set of grain arrangements compatible with fixed boundary load specifies an ensemble at fixed extensive stress ${\hat \sigma}V$ \cite{henkes07}. Similarly, the boundary of a packing's tiling can be constructed simply by knowing the boundary forces. If we then rearrange the grains or forces inside the packing to produce a new static packing compatible with the same boundary loading, the the tilings corresponding to the old and new packings will have the same area. Therefore the total area ${\cal A} = \sum_i a_i$ of the Maxwell-Cremona tilings in an ensemble at fixed extensive stress is {\em also} an additive invariant \cite{tighe08b,tighe09b}.  The invariance of $\cal A$ is a direct consequence of local force balance in the packing and has important consequences for the statistics of local stresses. 

\begin{figure}[tbp] 
\centering
\includegraphics[clip,width=0.65\linewidth]{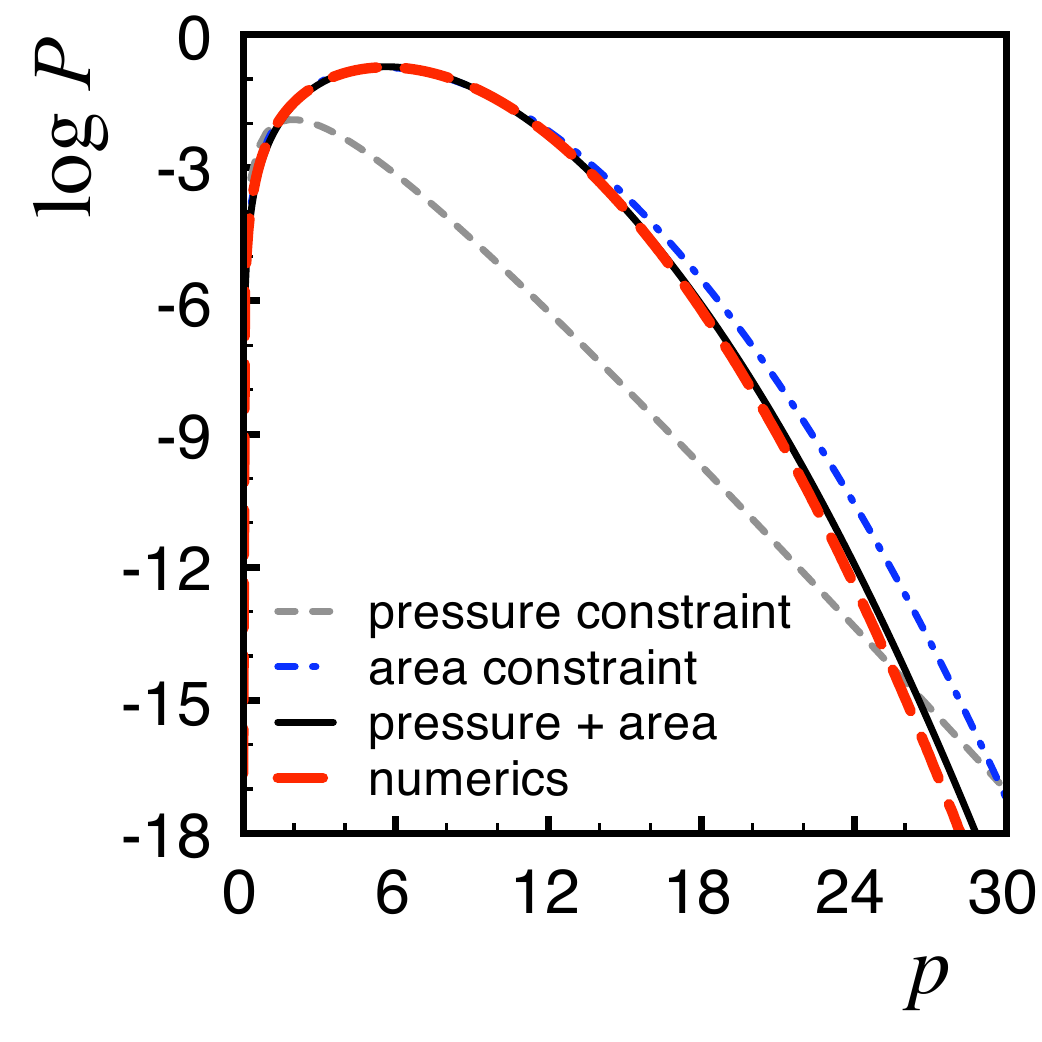}
\caption{ Numerical and theoretical probability distribution functions (see legend) of local pressure $P(p)$ in the  force network ensemble on a frictionless triangular lattice; adapted from Ref.~\cite{tighe08b}. The  numerical distribution is taken from a microcanonical ensemble of $N=1840$ grains using umbrella sampling \cite{vaneerd07,vaneerd09}. Theoretical distributions result from an entropy maximization calculation subject to a constraint on the average extensive pressure $\langle {\cal P} \rangle $ and/or average area of a reciprocal tile $\langle {\cal A} \rangle$. All calculations neglect correlations with neighboring grains, while the numerics impose  force balance exactly on every grain.}
\label{fig:naive}
\end{figure}

In Ref.~\cite{tighe08b} we considered an ideal gas-like calculation in which the average tiling area $\langle {\cal A} \rangle$, as well as the average extensive pressure $\langle {\cal P} \rangle$, is imposed. Maximizing entropy then leads to a local pressure distribution $P(p) = Z^{-1} p^\nu \exp{(-\alpha p - \gamma \langle a(p) \rangle )}$, where $\nu$ depends on contact geometry and $Z$, $\alpha$, and $\gamma$ are Lagrange multipliers. The distribution is asymptotically Gaussian because $\langle a(p) \rangle$, the average tile area given $p$, is proportional to $p^2$. As shown in Fig.~\ref{fig:naive}, this distribution is in excellent agreement with numerics in a model system, the force network ensemble of Snoeijer et al.~\cite{snoeijer04a}, and a clear improvement over an ideal gas-like calculation that does not enforce $\langle A \rangle$. As we discuss further below, the success of this approach is suggestive of a Boltzmann factor $\exp{(-\alpha {\cal P} - \gamma {\cal A})}$ that incorporates $\cal A$ in addition to $\cal P$.

The form of the tail of local stress distributions in static granular packings is a subject of ongoing debate, and our goal here is not to insist on one form over another. Rather, we ask what local stress distributions can tell us about applying the maximum entropy postulate to ensembles of packings. Here the force network ensemble is particularly useful, as it serves as a litmus test for theory. A calculation that is too simplistic to explain results in the force network ensemble {\em cannot} explain results in numerical or experimental ensembles, which are more complex than the force network ensemble; apparent agreement, if any, must be coincidental. Therefore, ideal gas-like calculations that predict exponential tails cannot explain the exponential tails observed in some experimental and numerical measurements \cite{radjai96,mueth98}, because the same calculations should apply to the force network ensemble, which does not display exponential tails. 

As the tiling area constraint is a necessary consequence of force balance and incorporating the constraint in an ideal gas-like calculation yields predictions consistent with the force network ensemble, it is important to ask how the tiling area should enter, more generally, in a maximum entropy approach.
A more detailed calculation than that in Ref.~\cite{tighe08b} is likely required to describe systems more complex than the force network ensemble, for example due to a growing correlation length \cite{wyart05,ellenbroek06}. 
The results of Ref.~\cite{tighe08b}, then, suggest two possible scenarios, but do not go far enough to distinguish between them. These are:
({\em i}) The extensive quantities $\cal P$ and $\cal A$ must be treated on equal footing, i.e.~the Boltzmann factor should in fact be $\exp{(-\alpha {\cal P} - \gamma {\cal A})}$. This possibility was already noted above. The second is that ({\em ii}) the tiling area $\cal A$ need not be enforced independently, that is it need not appear in the Boltzmann factor. Instead, its role is that of proxy for the spatial couplings implied by local force balance and neglected in an ideal gas calculation, and if these are incorporated exactly in an analytical calculation the tiling area $\cal A$ need not be considered separately. The goal of the present work is to distinguish between these two scenarios, which is a prerequisite for any future work that would seek to incorporate spatial correlations in more detail. The issue hinges fundamentally on the role of local force balance in a static granular ensemble. 

By considering two routes by which a microcanonical ensemble passes to a canonical ensemble, we argue that $\cal P$ and $\cal A$ are not independent in the thermodynamic limit, i.e.~we argue in favor of the latter of the two scenarios above. As support we offer the first analysis of the {\em canonical} force network ensemble. We will show that for small systems it matters a great deal which Boltzmann factor is used. For a thermodynamically large system, however, it suffices to characterize the system by either $\cal P$ or $\cal A$, provided one also imposes force balance {\em on every grain}. We confirm this through simulations of the force network ensemble, which allows us to access the canonical ensemble directly with Monte Carlo methods and to impose local force balance exactly. 

\section{ Force network ensemble}  The force network ensemble \cite{snoeijer04a} is an ideal testbed for statistical mechanics-based approaches to static granular media. We frame our discussion in the context of the force network ensemble, but our main conclusions regarding ensembles based on $\cal P$ and/or $\cal A$ apply to any ensemble of static granular packings. For now it suffices to state that, rather than comprising many different arrangements of the grains, the force network ensemble takes advantage of the fact that disk packings at finite pressure are generically {\em hyperstatic}, i.e.~the forces are underdetermined by the constraints of mechanical equilibrium. The ensemble then comprises all configurations of forces (force networks) on one quenched configuration of grains. In the microcanonical force network ensemble the global stress tensor is imposed, and all force balanced configurations of noncohesive forces are assigned equal statistical weight. For purposes of illustration we take the contact network to be a frictionless triangular lattice of grains and discuss at several points the relation to disordered packings. The force network ensemble is described in greater detail in Section \ref{sec:numerics}.

\section{ Passing to the canonical ensemble}  
\begin{figure}[tbp] 
\centering
\includegraphics[clip,width=0.8\linewidth]{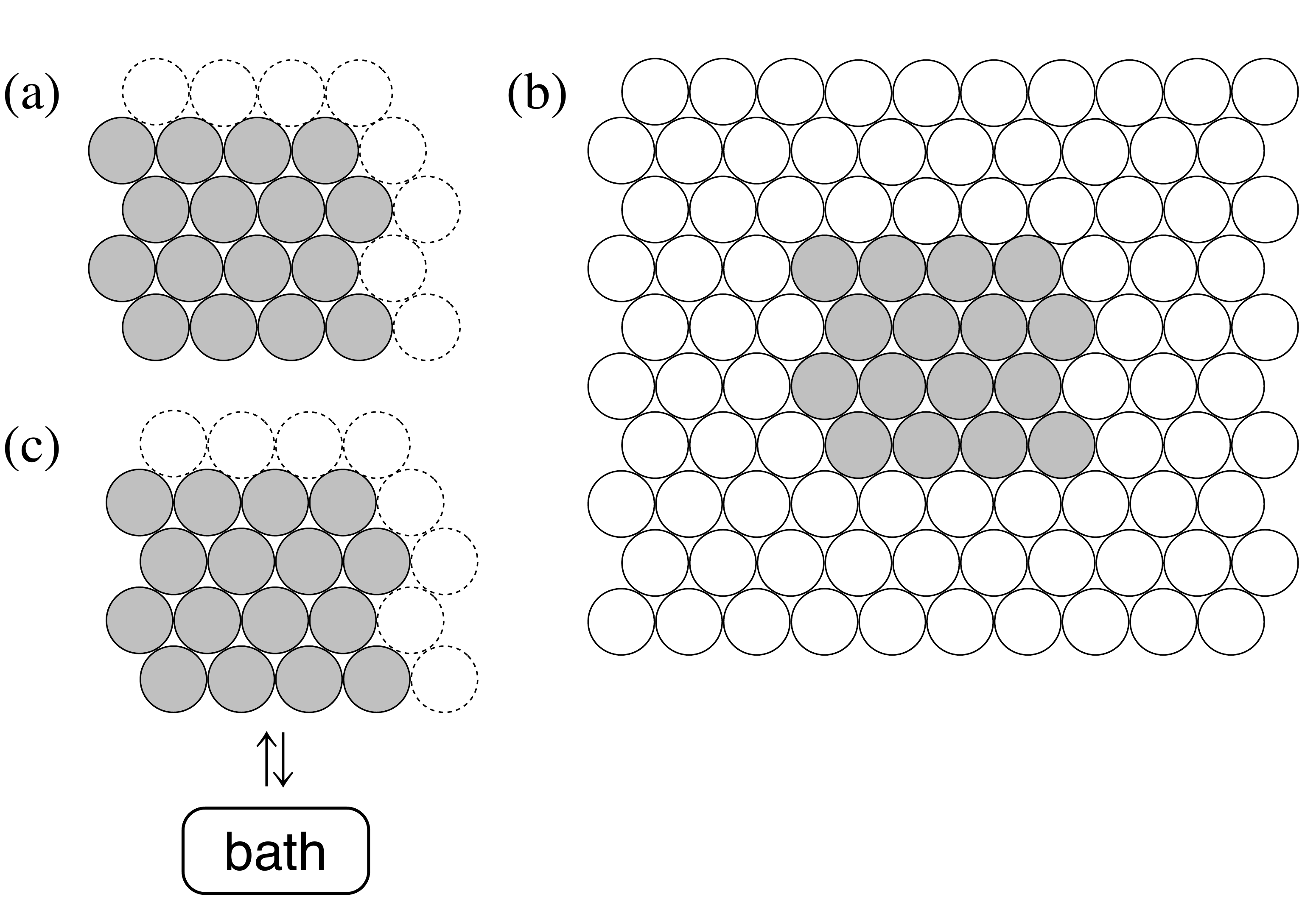}
\caption{(a) A periodic triangular lattice. In the force network ensemble, the set of force networks (see Fig.~\ref{fig:tiling}a) with fixed stress tensor comprise the microcanonical force network ensemble on this fixed contact network. (b) A canonical ensemble can be created by embedding the system of (a) in a larger packing that then acts as a bath. The extensive stress of the system and bath, together, is conserved, and the canonical system is no longer periodic. (c) Alternatively, the system of (a) can be placed in contact with a bath, i.e.~allowed to exchange extensive stress with the bath in such a way that the extensive stress of the two systems is conserved {\em and} the canonical system remains periodic.}
\label{fig:canonical}
\end{figure}

As a thought experiment, one passes from a microcanonical to a canonical ensemble by placing a previously microcanonical system in contact with a much larger system that acts as a reservoir of some conserved quantity, e.g.~$\cal P$. For ensembles of static packings, this can be achieved by sampling $N$-grain clusters of grains from a microcanonical ensemble of packings of $M \gg N$ grains. This is illustrated in Fig.~\ref{fig:canonical}a and b. The extensive pressure in the canonical system ${\cal P}_\nu$ fluctuates from cluster to cluster, but its average is dictated by the bath: $\langle {\cal P}_\nu \rangle = N\langle p \rangle$, where $\langle p \rangle$ is the average pressure per grain in the bath. Formally, the parameter $\alpha$ in the Boltzmann factor $\exp{(-\alpha {\cal P}_\nu)}$ plays the role of a Lagrange multiplier that imposes this constraint on the canonical system.

The twist is that, as noted above, a microcanonical system at fixed $\cal P$ also has fixed tiling area $\cal A$. Therefore the bath also imposes an average tiling area $\langle {\cal A}_\nu \rangle = N\langle a \rangle$ on the canonical system, where $\langle a \rangle$ is the average area per tile in the bath. If the imposed average tiling area is independent of the imposed average extensive pressure, then an additional Lagrange multiplier $\gamma$ is needed, i.e.~the Boltzmann factor should be $\exp{(-\alpha {\cal P}_\nu - \gamma{\cal A}_\nu)}$. 

\begin{figure}[tbp] 
\centering
\includegraphics[clip,width=0.8\linewidth]{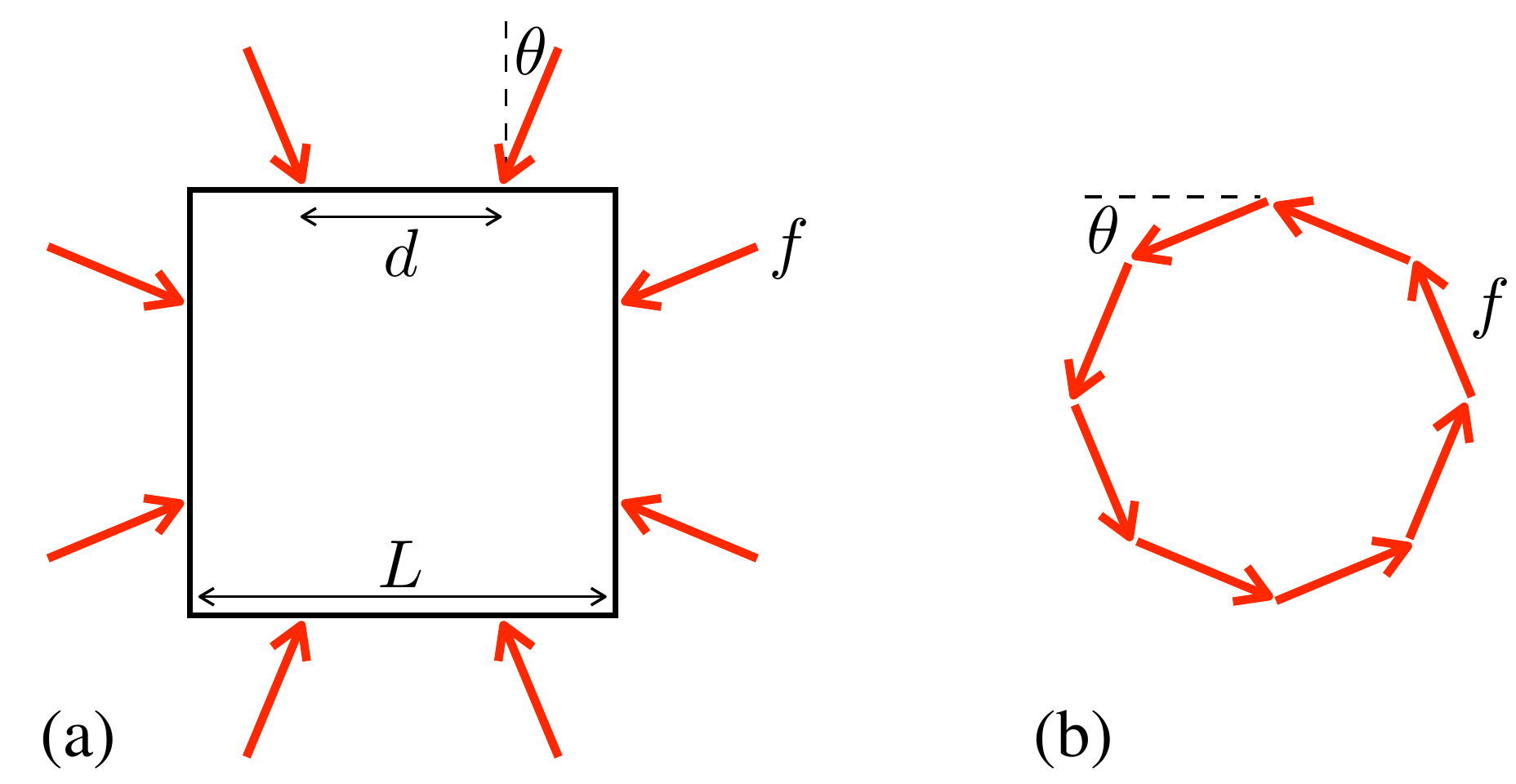}
\caption{ (a) Boundary loading of a square body. (b) The boundary of the reciprocal tiling of the system in (a). With the appropriate variation of boundary force magnitude $f$, orientation $\theta$, and separation $d$, the area of the tiling can be changed while the stress is held fixed. }
\label{fig:nonperiodic}
\end{figure}

To illustrate the potential complications introduced by $\cal A$ in a canonical ensemble, consider the non-periodic canonical system depicted in Fig.~\ref{fig:nonperiodic}. (We now dispense with the subscript $\nu$ for extensive quantities in canonical systems.) The system with dimensions $L \times L$ experiences a loading due to forces imposed by the bath. In terms of its boundary forces $\lbrace {\vec f}_c \rbrace$ acting at positions $\lbrace {\vec x}_c \rbrace$, the system's extensive stress is \cite{landau} 
\footnote{It can be shown that Eq.~(\ref{eqn:avgstress}) is equivalent to the more familiar expression $\sigma_{\alpha \beta}V = \frac{1}{2}\sum_{ij} f_{ij,\alpha}\,r_{ij,\beta}$.
}
\begin{equation} \label{eqn:avgstress}
\sigma_{\alpha \beta}V = \sum_c f_{c,\alpha}\,x_{c,\beta} \,.
\end{equation}
For boundary forces of magnitude $f$ at orientation $\theta$, separated by a distance $d$ as in Fig.~\ref{fig:nonperiodic}, the system's extensive stress tensor is isotropic, with extensive pressure ${\cal P} = 2f(L\cos{\theta}+d\sin{\theta})$. The tiling area is ${\cal A} = 2f^2(1 + \sin{2\theta}+\cos{2\theta})$. Note that  $\cal A$ can be changed while holding $\cal P$ fixed by varying $f$ and $\theta$  and requiring $d = ({\cal P} -2fL\cos{\theta})/2f\sin{\theta}$. 

The above example suffices to show that ${\cal A}$ cannot be a single-valued function of ${\cal P}$ in a non-periodic system. Introducing more, and more disordered, boundary forces will only enhance this degeneracy, as there will be many ways to choose the boundary forces to achieve a particular $\cal P$, and these configurations will have different tiling areas. 
The question is whether the relative area fluctuations $\sqrt{\langle \delta {\cal A}^2 \rangle}/\langle {\cal A}\rangle$ at fixed $\cal P$ become negligible in the thermodynamic limit; if so, imposing $\langle {\cal P} \rangle$ via the intensive parameter $\alpha$ also suffices to select $\langle {\cal A} \rangle$. If not, an additional thermodynamic parameter conjugate to $\cal A$ must be introduced.

\subsection{Periodic canonical systems}
Before returning to the fluctuations of tiling area in the system of Fig.~\ref{fig:canonical}b, it is useful to consider an alternate route from the microcanonical to the canonical ensemble. While in Fig.~\ref{fig:canonical}b the canonical system is non-periodic and embedded in a larger system, we now imagine placing a {\em periodic} system in contact with a reservoir with which it is allowed to exchange extensive stress, as illustrated in Fig.~\ref{fig:canonical}c. Unlike the previous scenario, which can be understood as clusters sampled from a larger microcanonical ensemble, this scenario 
corresponds to a thought experiment in which a collection of $N$ grains is repeatedly randomly seeded inside a prescribed unit cell and allowed to relax to mechanical equilibrium. Final states that are mechanically stable are placed in the ensemble and weighted according to the Boltzmann factor. 

The key observation is that, in a periodic force network, the tiling area is a single-valued function of the extensive stress tensor: ${\cal A} = ({\rm det}\,{\hat \sigma})V$. To see this consider the force network and tiling in Fig.~\ref{fig:tiling}a and c. Imposing the stress tensor is equivalent to imposing the net vector forces ${\vec F}_1$ and ${\vec F}_2$ acting on two boundaries of the $L_1 \times L_2$ unit cell. For the force network in Fig.~\ref{fig:tiling}a the extensive stress tensor is
\begin{equation}
\hat{\sigma} V = 
\left(
\begin{array}{cc}
F_1 L_1 & 0 \\
0 & F_2 L_2
\end{array}
\right) \,
\end{equation}
where $V = L_1 L_2$. It is straightforward to see that, due to periodicity, the area of the network's reciprocal tiling must be ${\cal A} = F_1 F_2 =({\rm det}\,{\hat \sigma})V$. As one can always construct a rectangular unit cell and choose coordinate axes aligned with the principal stress directions, the relation ${\cal A} = ({\rm det}\,{\hat \sigma})V$ is general for periodic force balanced networks in two dimensions, independent of contact network geometry and topology.

The relation between tiling area and stress has immediate consequences. First, in a canonical ensemble of periodic packings (Fig.~\ref{fig:canonical}c) it is clear that extensive pressure $\cal P$ and tiling area $\cal A$ are not independent; knowledge of $\cal P$ implies knowledge of $\cal A$. For ensembles of noncohesive isotropic packings, the relation is bidirectional: $\cal P$ can be inferred from $\cal A$.  Therefore it suffices to employ a Boltzmann factor $\exp{(-\alpha {\cal P})}$ or, alternatively, $\exp{(-\gamma {\cal A})}$. This latter possibility has not previously been considered, and we test it below. 

The second consequence follows from noting that the routes to the canonical ensemble depicted in Fig.~\ref{fig:canonical}b and c differ only in their implementation of boundary conditions. In general, one anticipates that details at the boundary should not influence the thermodynamic limit, and thus in the thermodynamic limit a Boltzmann factor $\exp{(-\alpha {\cal P})}$ or $\exp{(-\gamma {\cal A})}$ will also suffice for non-periodic systems. Consequently, it is not necessary to employ a Boltzmann factor $\exp{(-\alpha {\cal P} -\gamma {\cal A})}$ in a thermodynamically large system. We reinforce this expectation below by providing a scaling argument for the relative area fluctuations in a non-periodic system. We will test these predictions by performing simulations of the canonical force network ensemble for both periodic and non-periodic force networks of varying size $N$. We emphasize that equivalence in the thermodynamic limit does not imply equivalence in small systems, a point we discuss further below.

\subsection{Non-periodic canonical systems}
Returning to non-periodic canonical systems, as in Fig.~\ref{fig:canonical}b, we consider the tiling area fluctuations in the thermodynamic limit. We argue that for each extensive pressure $\cal P$ the system approaches the same tiling area selected in periodic systems with vanishing relative fluctuations.

\begin{figure}[tbp] 
\centering
\includegraphics[clip,width=0.7\linewidth]{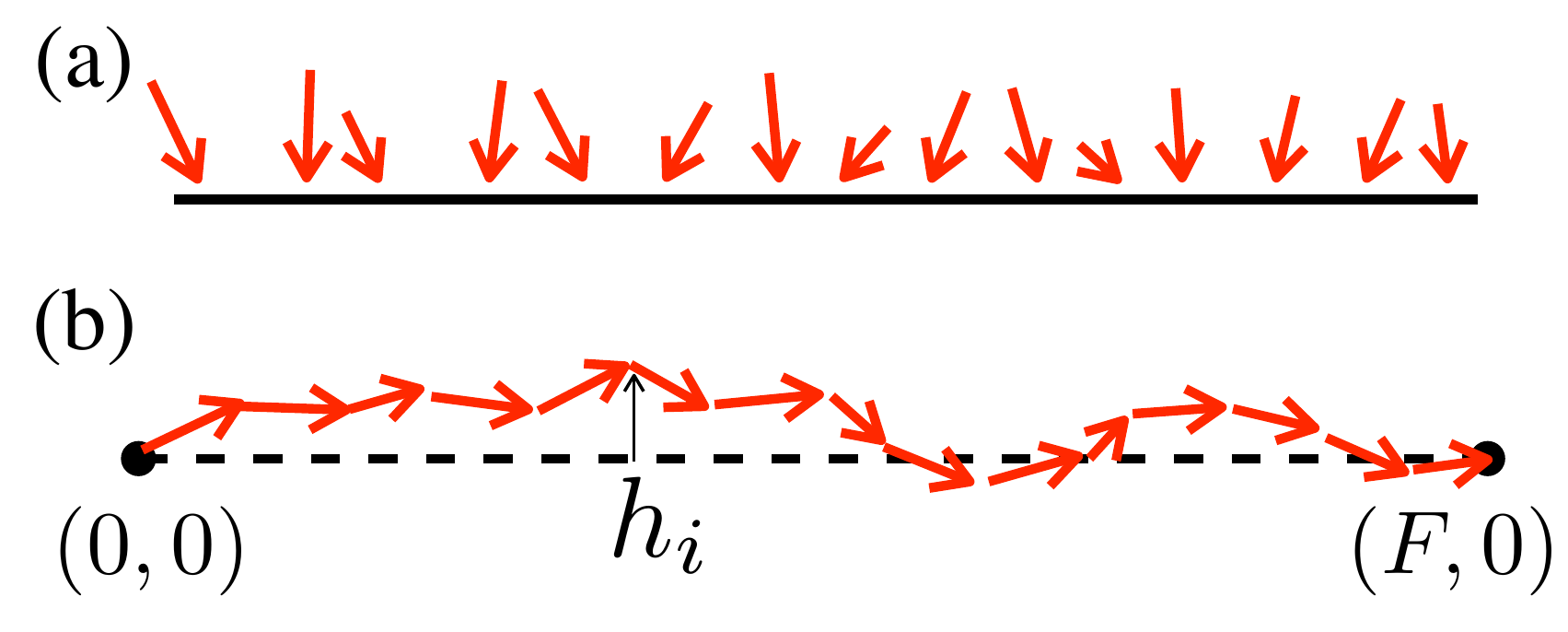}
\caption{(a) Boundary of a non-periodic structure subject to externally imposed forces. The net force $(-F,0)$ is purely compressive. (b) The boundary of the structures reciprocal tiling can be imposed from its boundary forces. It resembles a directed random walk constrained to travel a distance $(F,0)$. Its vertical displacement after $i$ steps is labeled $h_i$. }
\label{fig:randomwalk}
\end{figure}

Consider a square system subject to isotropic compressive loading. One of its boundaries is depicted in Fig.~\ref{fig:randomwalk}a. The boundary is subject to a net force $(0,-{ F})$, with ${ F} \simeq {\cal P}/L \sim O(\sqrt{N})$. The $N_{\rm b} \sim O(\sqrt{N})$ boundary forces can be used to construct one boundary of the system's reciprocal tiling, shown in Fig.~\ref{fig:randomwalk}b. This boundary resembles a directed random walk starting from the origin and constrained to end at $(F,0)$. Each possible walk is as likely as its reflection about the dashed segment in Fig.~\ref{fig:randomwalk}b. It follows that the mean tiling area will be $\langle {\cal A} \rangle = F^2 = ({\rm det}\,{\hat \sigma})V$, just as in a periodic system.

It remains to consider the dependence on $N$ of the tiling area fluctuations $\sqrt{\langle \delta {\cal A}^2 \rangle}$. We assume each step of the walk goes a typical distance $\langle f \rangle = F/N_{\rm b}$ to the right in Fig.~\ref{fig:randomwalk}b. In the following the load per unit length $\langle f \rangle = F/N_{\rm b}$ is kept constant as the system size increases. Labeling $h_i$ the $y$-coordinate of the walker at step $i$, the typical vertical distance traveled after $i$ steps is $\sqrt{\langle h_i^2 \rangle}$. Sufficiently close to the left endpoint of the walk, the influence of the constraint to end at $(F,0)$ will not be felt, and we must have $\sqrt{\langle h_i^2 \rangle} \simeq \langle f \rangle \sqrt{i}$. Similarly, near the right endpoint $ \sqrt{\langle h_i^2 \rangle} \simeq \langle f \rangle \sqrt{N_{\rm b} - i}$.

We want to develop an upper bound on the tiling area fluctuations. In this spirit, we assume that the typical vertical displacement behaves like a simple random walk up to the midpoint, i.e.
\begin{equation}
\sqrt{\langle h_i^2 \rangle} \simeq 
\left \lbrace
\begin{array}{ll}
\langle f \rangle \sqrt{i} 				& i< \frac{1}{2}N_{\rm b} \\
\langle f \rangle \sqrt{N_{\rm b} - i} 	& i> \frac{1}{2}N_{\rm b}  \,.
\end{array} 
\right.
\label{eqn:randomwalk}
\end{equation}
The scaling of absolute area fluctuations under a simple random walk is captured by $\sqrt{\langle \delta {\cal A}^2 \rangle} \simeq \langle f \rangle  \sum_i \sqrt{ \langle h_i^2 \rangle}$.
Using (\ref{eqn:randomwalk}) we find 
$\sqrt{\langle \delta {\cal A}^2 \rangle} \lesssim \langle f \rangle^2 N_{\rm b}^\frac{3}{2} \sim O(N^\frac{3}{4})$.
As $\langle {\cal A} \rangle$ is extensive, the relative fluctuations $\sqrt{\langle \delta {\cal A}^2 \rangle}/\langle {\cal A}\rangle$ decay at least as fast as $N^{-\frac{1}{4}}$, and therefore vanish in the thermodynamic limit.

Note that our argument relies on treating boundary forces as random variables. For a  packing at isostaticity this cannot hold; specifying only half the boundary forces of an isostatic packing suffices to fix the other half \cite{tkachenko99}. We now show that  if a packing has mean contact number $z = z_{\rm iso} + \Delta z$, there is a length scale $\ell^\star$ such that, for packings of linear dimension $L > \ell^\star$, all the boundary forces can indeed be treated as random variables and the above scaling argument can again be invoked. In order to randomly assign all the $N_{\rm b}$ boundary forces, there must be enough underdetermined forces in the bulk.  There are $N_{\rm e} \simeq \Delta z \, N$ of these ``excess'' forces. For $N_{\rm e}$ to balance or exceed $N_{\rm b}$ requires $\Delta z \gtrsim N^{-\frac{1}{2}}$. As the system's linear dimension $L \sim \sqrt{N}$, this is equivalent to requiring $L \gtrsim 1/\Delta z$. Note that the same balance between boundary contacts and excess bulk contacts is also used to derive the isostatic length $\ell^\star \sim 1/\Delta z$ \cite{wyart05,ellenbroek06}, which governs the crossover from discrete to continuum response in static packings. Therefore our argument predicts that relative area fluctuations vanish in the thermodynamic limit for systems arbitrarily close to isostaticity, provided the system size $L$ exceeds the diverging length scale $\ell^\star \sim 1/\Delta z$. Our scaling argument cannot be applied to systems smaller than $\ell^\star$, so it may be possible to identify stronger bounds on the area fluctuations.

\section{ Numerical results}  \label{sec:numerics}

\subsection{Monte Carlo methods}
The force network ensemble is a convenient venue to test the ideas described above. Because the ensemble can be sampled with Monte Carlo methods, it is possible to realize, numerically, the thought experiments described in Fig.~\ref{fig:canonical}. 

We briefly summarize the force network ensemble and the Monte Carlo methods we employ to sample the canonical ensemble. Having selected a hyperstatic contact network, the key requirement is the imposition of force balance on each grain $i$, i.e.~$\sum_j {\vec f}_{ij} = 0$, where ${\vec f}_{ij}$ is the contact force on $i$ due to $j$. The microcanonical force network ensemble also imposes the stress $\sigma_{\alpha \beta} = (r/V)\sum_{ij}f_{ij} n_{ij, \alpha} n_{ij,\beta}$, where $r$ is the grain radius, $V$ is the area of the quenched packing, and ${\vec n}_{ij}$ is the unit vector from $i$ to $j$. We restrict ourselves to isotropic systems, and the important quantity will be the extensive pressure ${\cal P} = (V/2) {\rm Tr}\, {\hat \sigma} = (r/2) \sum_{ij} f_{ij}$. We also impose a positivity condition on each force, $f_{ij} \ge 0$, restricting the ensemble to noncohesive force networks. Any configuration of forces ${\bf f} = \lbrace f_{ij} \rbrace$ satisfying these constraints is an element of the microcanonical force network ensemble; Fig.~\ref{fig:tiling}a gives an example. 

The ensemble can be efficiently and flatly sampled using Monte Carlo techniques. For periodic systems we use the ``wheel moves'' of Ref.~\cite{tighe05}. These may be modified to work in non-periodic systems, as well \cite{vaneerd07,vaneerd09}. Changes to boundary forces are allowed provided they respect force balance and noncohesiveness of the forces. In the canonical ensemble we add an additional move that changes the extensive pressure $\cal P$. In a triangular lattice, this move simply adds a quantity $\epsilon$ to each force. A proposed move from a force network ${\bf f}_\nu$ to a force network ${\bf f}_{\nu^\prime}$ is accepted with a probability determined according to the Metropolis acceptance rule,
\begin{equation}
{\rm acc}({\bf f}_{ \nu} \rightarrow {\bf f}_{ \nu^\prime})
 = {\rm min}\left( 
 1, \frac{B({\bf f}_{\nu^\prime})}{B({\bf f}_{\nu})}
 \right)\,\Theta({\bf f}_{\nu^\prime}) \,.
\end{equation}
Here $B({\bf f}_\nu)$ is the Boltzmann factor, which we vary below. $\Theta({\bf f}_{\nu^\prime})=1$ if the configuration ${\bf f}_{\nu^\prime}$ is force balanced and noncohesive and 0 otherwise. 

\subsection{Microcanonical ensemble}
\begin{figure}[tbp] 
\centering
\includegraphics[clip,width=0.5\linewidth]{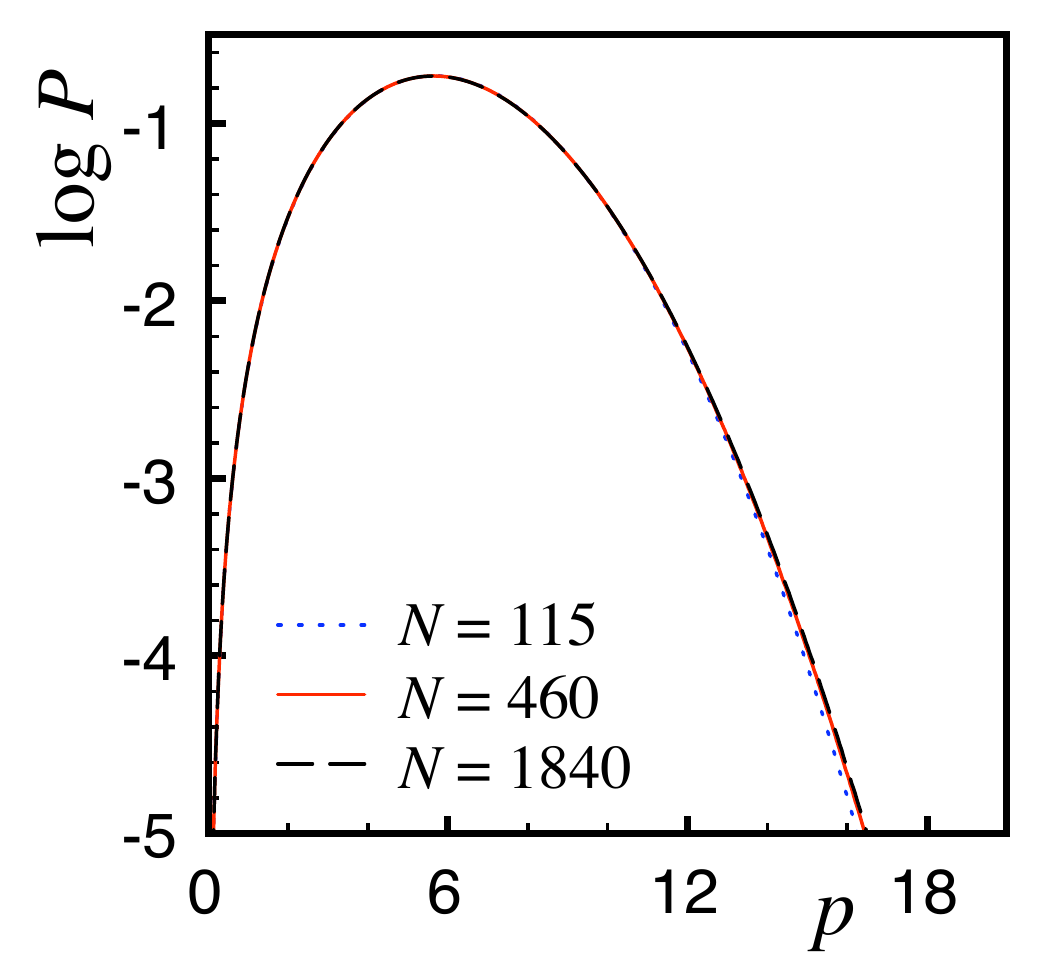}
\caption{ Local pressure probability distribution $P(p)$ in the microcanonical force network ensemble on a periodic frictionless triangular lattice for unit cells containing $N=115$, 460, and 1840 grains. Finite size effects are negligible. }
\label{fig:micro}
\end{figure}

All previous simulations of the force network ensemble have been of a microcanonical variety \cite{snoeijer04b,tighe05,tighe08b,tighe09b,snoeijer04a,vaneerd07,vaneerd09}. The microcanonical frictionless triangular lattice has been well-studied in the force network ensemble, and local stress statistics have been determined extremely accurately by employing umbrella sampling \cite{tighe08b,vaneerd07}. Fig.~\ref{fig:micro} plots the microcanonical $P(p)$ for systems of $N = 115$, 460, and 1840 grains, demonstrating that finite size effects are negligible in the largest system. 

\subsection{Canonical ensembles}
To test the above predictions we perform simulations of the canonical force network ensemble on periodic and non-periodic triangular lattices. We consider two ensembles equivalent if their local pressure distributions $P(p)$ converge to the same limiting distribution as $N \rightarrow \infty$. We take advantage of the fact that $P(p)$ is known extremely accurately in the microcanonical ensemble (Fig.~\ref{fig:micro}) and consider $P(p)$ in a microcanonical ensemble of $N = 1840$ grains representative of the thermodynamic limit. By studying canonical systems of varying system size $N$, we look for numerical evidence of convergence to $P(p)$ from the large microcanonical system. By checking if different canonical ensembles converge to this distribution, we can compare them to each other by transitivity. We emphasize the perspective that $\alpha$ and $\gamma$ are Lagrange multipliers: for systems of different size $N$, thermodynamic parameters are not assigned but varied in order to achieve a predetermined $\langle {\cal P} \rangle/N$ and/or $\langle {\cal A} \rangle/N$.

\begin{figure}[tbp] 
\centering
\includegraphics[clip,width=0.49\linewidth]{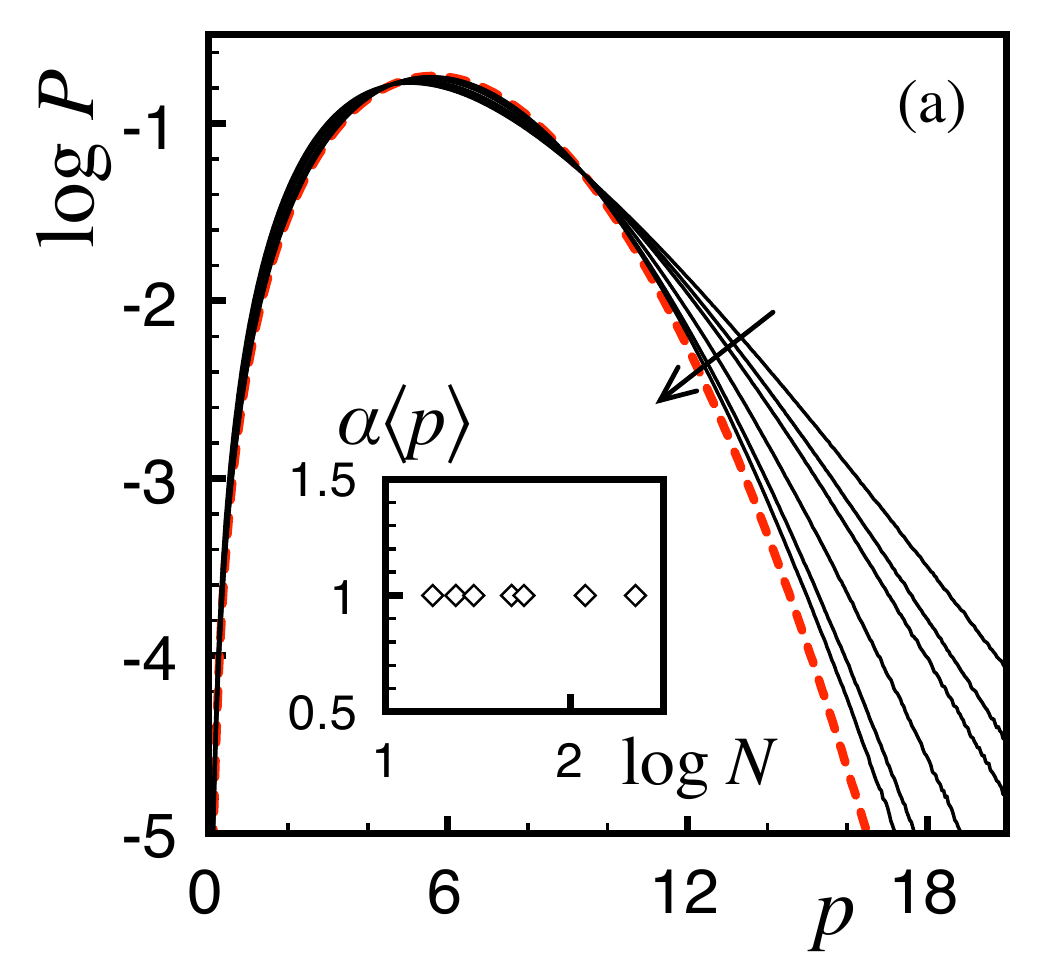}
\includegraphics[clip,width=0.49\linewidth]{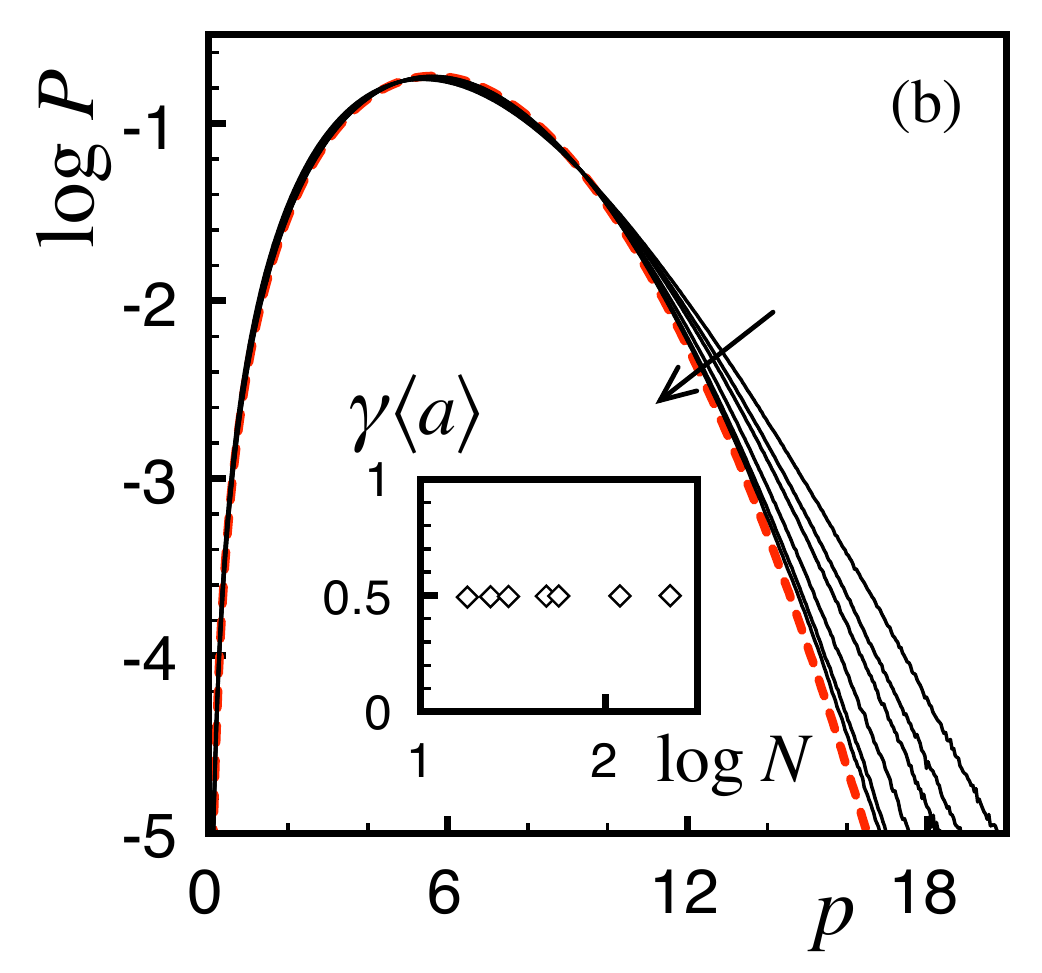}
\caption{(a) Local pressure probability distribution $P(p)$ in the canonical force network ensemble in periodic systems with Boltzmann factor $\exp{(-\alpha{\cal P})}$ (solid curves) and a microcanonical system with $N = 1840$ (dashed curve). The canonical system size $N=18$, 24, 30, 56, 120, and 224 increases in the direction of the arrow. (inset) The parameter $\alpha$ is insensitive to system size and consistent with $\alpha \langle p \rangle=1$. (b) $P(p)$ in the canonical ensemble in periodic systems with Boltzmann factor $\exp{(-\gamma{\cal A})}$ (solid curves) and microcanonical ensemble (dashed curve). System sizes are identical to (a). (inset) The parameter $\gamma$ is insensitive to system size and consistent with $\gamma\langle a \rangle = \frac{1}{2}$. }
\label{fig:numerics1}
\end{figure}

We first test the prediction that a periodic canonical ensemble can be described using either of two Boltzmann factors. Fig.~\ref{fig:numerics1}a depicts the case $B({\bf f}_{\nu}) = \exp{(-\alpha {\cal P}_\nu)}$. Systems of varying size $N$ are simulated, and $\alpha$ is chosen so that $\langle {\cal P} \rangle = 6N$. The probability distribution $P(p)$ of the local pressure $p$ is sampled, and converges to the distribution sampled from a large microcanonical force network ensemble. $\alpha$ is nearly constant over the range of system sizes sampled and consistent with its expected value in the thermodynamic limit $\alpha \langle p \rangle = 1$ (see Appendix). This strongly indicates that the microcanonical force network ensemble and the canonical force network ensemble with Boltzmann factor $\exp{(-\alpha {\cal P})}$ are equivalent in the thermodynamic limit. It is remarkable that for small systems the asymptotic decay of $P(p)$ appears nearly exponential, though this is not the behavior in the thermodynamic limit. As always, caution is required when interpreting simulations or measurements in small systems.

Fig.~\ref{fig:numerics1}b depicts the case $B({\bf f}_{\nu}) = \exp{(-\gamma {\cal A}_\nu)}$ applied to an ensemble of periodic force networks. The probability distribution $P(p)$ is again sampled for varying system size. The intensive parameter $\gamma$ is chosen to ensure $\langle A \rangle =   \frac{3\sqrt{3}}{2}N$, which follows from ${\cal A} = ({\rm det}\,{\hat \sigma})V$ for $\langle {\cal P} \rangle = 6N$. The pressure statistics again converge to the microcanonical distribution in Fig.~\ref{fig:micro}, and the thermodynamic parameter $\gamma$ is independent of $N$ and consistent with its expected value $\gamma \langle a \rangle = \frac{1}{2}$ (see Appendix). This provides strong evidence that the microcanonical force network ensemble and the canonical force network ensemble with Boltzmann factor $\exp{(-\gamma {\cal A})}$ are equivalent in the thermodynamic limit. By transitivity it also indicates equivalence of the two canonical ensembles in the thermodynamic limit.

\begin{figure}[tbp]  
\centering
\includegraphics[clip,width=0.49\linewidth]{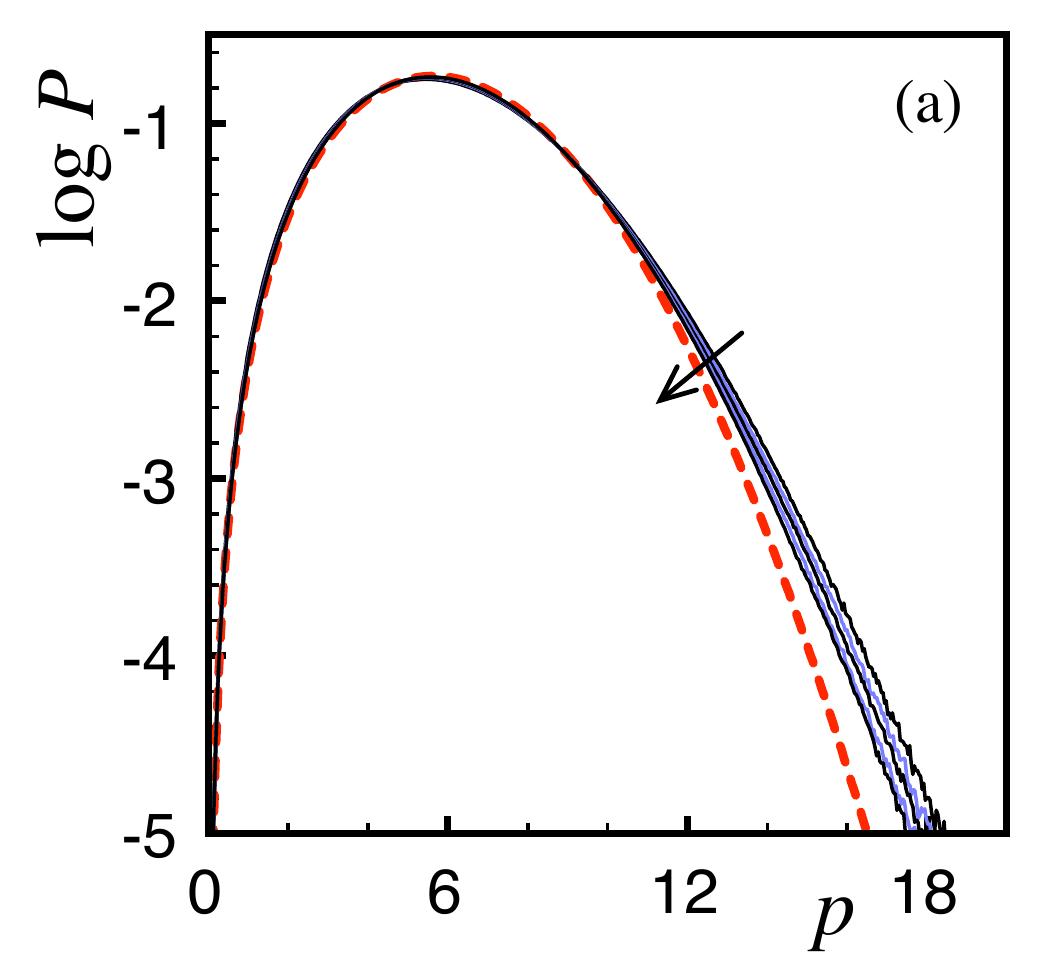}
\includegraphics[clip,width=0.49\linewidth]{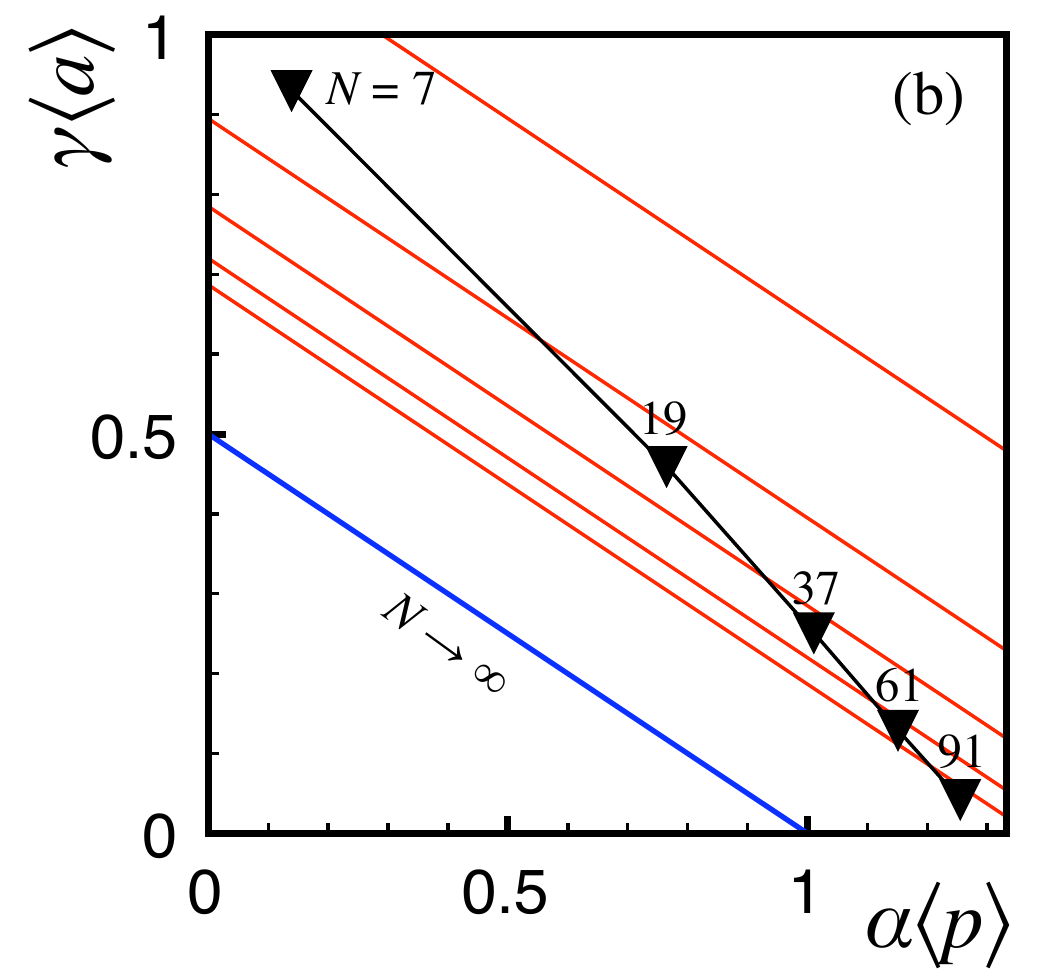}
\caption{(a) Local pressure probability distribution $P(p)$ in the canonical force network ensemble for non-periodic systems with Boltzmann factor $\exp{(-\alpha{\cal P} - \gamma{\cal A})}$ (solid curves) and a periodic microcanonical system (dashed curve). The canonical system size ($N=7$, 19, 37, 61, and 91, corresponding to 1, 2, 3, 4, and 5 hexagonal layers around a central grain, respectively) increases in the direction of the arrow. The microcanonical distribution is identical to that in Figs.~\ref{fig:naive} and \ref{fig:numerics1}. (b) Evolution of the parameters $\alpha$ and $\gamma$ (triangles) with system size. Lines are contours of  (\ref{eqn:approx}) for the same systems sizes, as well as the thermodynamic limit.  }
\label{fig:convergence}
\end{figure}

We next consider non-periodic systems. Packings are composed of grains organized in hexagonal layers around a central grain. $P(p)$ is sampled on the central grain. We employ a Boltzmann factor $B({\bf f}_{\nu}) = \exp{(-\alpha {\cal P} - \gamma {\cal A})}$ and choose $\alpha$ and $\gamma$ such that $\langle {\cal P} \rangle = 6N$ and $\langle {\cal A} \rangle = \frac{3\sqrt{3}}{2}N$. If the two constraints are not independent, one of the thermodynamic parameters must tend to zero. The thermodynamic parameters $\alpha$ and $\gamma$ are plotted in Fig.~\ref{fig:convergence}b. Indeed, $\gamma$ tends to zero with increasing system size, consistent with argument that relative tiling area fluctuations vanish with increasing $N$. 

Though $\alpha$ in Fig.~\ref{fig:convergence}b exceeds its value from periodic packings (Fig.~\ref{fig:numerics1}a inset), we argue that it will ultimately approach the same limiting value. It may be shown (see Appendix) that the two parameters satisfy the relation
\begin{equation} \label{eqn:approx}
\alpha \langle p \rangle + 2\gamma \langle a \rangle \approx \frac{N_{\rm c}}{N} - 2 \,,
\end{equation}
where $N_{\rm c}$ is the number of contact forces in the system. Contours of   (\ref{eqn:approx}) are plotted in Fig.~\ref{fig:convergence}. The approximations implicit in  (\ref{eqn:approx}) become increasingly accurate as the system size increases; indeed, the largest system sizes in Fig.~\ref{fig:convergence} lie on their respective contours. The intercept value $\frac{N_{\rm c}}{N} - 2$ of the contours provides an upper bound on $\alpha \langle p \rangle$ and smoothly approaches its limiting value $\frac{N_{\rm c}}{N} - 2 \rightarrow \frac{1}{2}(z-z_{\rm iso})$ from above, where $z=6$ is the mean contact number and $z_{\rm iso}$ is the isostatic coordination number for frictionless disk packings. Hence, although $\alpha \langle p \rangle$ overshoots its asymptotic limit, we anticipate that it will approach the asymptotic value smoothly for system sizes larger than we can simulate practically.

\begin{figure}[tbp] 
\centering
\includegraphics[clip,width=0.49\linewidth]{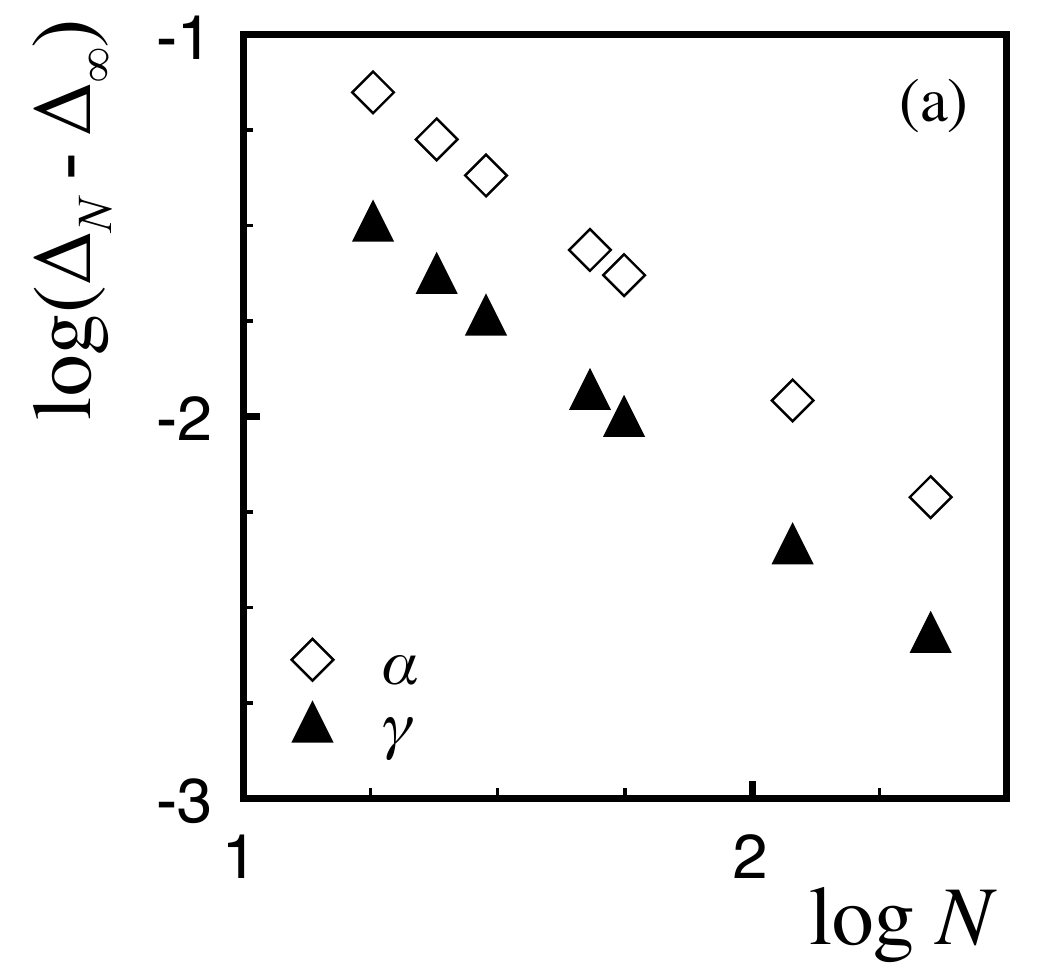}
\includegraphics[clip,width=0.49\linewidth]{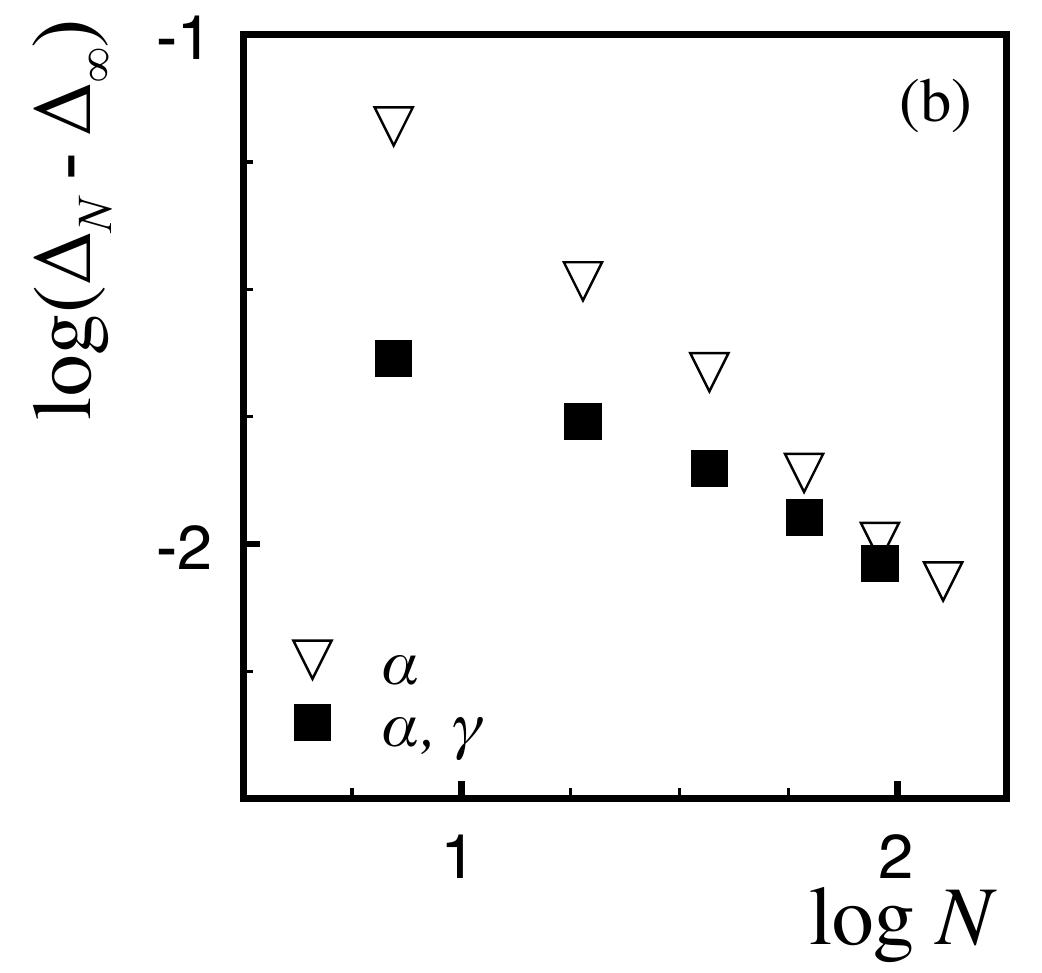}
\caption{Evolution of the relative fluctuations $\Delta_N$with system size. Fluctuations from finite size systems are taken from (a) the periodic systems of Fig.~\ref{fig:numerics1} with Boltzmann factors $\exp{(-\alpha{\cal P})}$ and  $\exp{(-\gamma{\cal A})}$, and (b) the non-periodic systems of Fig.~\ref{fig:convergence} with Boltzmann factor $\exp{(-\alpha{\cal P} - \gamma{\cal A})}$, as well as non-periodic systems sampled with the Boltzmann factor $\exp{(-\alpha{\cal P})}$ (distributions of $P(p)$ not shown). The fluctuations in the thermodynamic limit are estimated from the distribution in Fig.~\ref{fig:naive}. }
\label{fig:relfluct}
\end{figure}

It is evident from Figs.~\ref{fig:numerics1} and \ref{fig:convergence}a that local pressure statistics in small systems are closer to their asymptotic form in the periodic system with Boltzmann factor $\exp{(-\gamma {\cal A})}$ than in the periodic system with Boltzmann factor $\exp{(-\alpha {\cal P})}$. For the smallest systems they are comparatively closest to their asymptotic form in the non-periodic system with Boltzmann factor $\exp{(-\alpha{\cal P} -\gamma {\cal A})}$. This is quantified in Fig.~\ref{fig:relfluct}, which plots the convergence of local pressure fluctuations $\Delta_N := \sqrt{\langle \delta p^2 \rangle_N} / \langle p \rangle_N $ to their $N \rightarrow \infty$ value, approximated by $\Delta_N$ evaluated for the numerically sampled microcanonical distribution for $N = 1840$. For all cases, $\Delta_N$ decays with $N$, confirming the convergence of canonical and microcanonical ensembles that was already apparent from Figs.~\ref{fig:numerics1} and \ref{fig:convergence}. For the system sizes sampled in both periodic and non-periodic systems, $\Delta_N$ is always smaller when the intensive thermodynamic parameter $\gamma$ is employed. In periodic systems $(\Delta_N - \Delta_\infty) \simeq N^{-1}$, while in non-periodic systems the decay with system size is weaker. 

\section{Conclusions}
Static packings possess reciprocal tilings as a consequence of local force balance. The area of the tiling is therefore an additive extensive quantity that is {\em induced} by the presence of local force balance. We have considered the role of this tiling area, and through it force balance, in the construction of a canonical ensemble of static granular packings.

In a canonical ensemble both the extensive stress and the tiling area fluctuate from configuration to configuration. These fluctuations are not independent: viewing the bath as a microcanonical system in which the canonical (sub)system is embedded, the bath pressure ${\cal P} = \frac{1}{2}({\rm Tr}\,{\hat \sigma})V$ and tiling area ${\cal A} = ({\rm det}\,{\hat \sigma})V$ are related. Imposing an average extensive pressure $\langle {\cal P} \rangle$ in the canonical system, via a conjugate intensive parameter $\alpha$, also suffices to select the proper average tiling area $\langle {\cal A} \rangle$ in the thermodynamic limit. This expectation was confirmed both with a scaling argument for the relative tiling area fluctuations and numerical simulation. 

Note that our simulations directly addressed only ordered packings far from isostaticity. Recalling that the relation between $\cal A$ and $\cal P$ in periodic systems holds regardless of disorder, there is no reason to expect disorder to prevent convergence in the thermodynamic limit. A disordered system must of course be large enough to average over fluctuations in the local contact number; therefore, {\em ceteris paribus}, a numerical measure of convergence such as $\Delta_N$ will have a smaller prefactor for ensembles of ordered packings than for disordered packings. We argued above that in the limit $\Delta z \downarrow 0$, convergence to the thermodynamic limit can only be expected for systems of size $L \gg \ell^\star \sim 1/\Delta z$. This is consistent with previous studies \cite{ellenbroek06} but should be tested explicitly. Numerical tools beyond the force network ensemble are required, however, because the ensemble itself vanishes in the isostatic limit. 

We have also demonstrated that a Boltzmann factor $\exp{(-\gamma {\cal A})}$ may be used in place of $\exp{(-\alpha {\cal P})}$. This may seem a curiosity; e.g.~ the equilibrium Boltzmann factor $\exp{(-\beta E)}$ can also be replaced with $\exp{(-\tilde{\beta} E^2)}$ in a purely repulsive system, though additivity is lost and no insight is gained. There are important differences, however, for ensembles of static packings. Unlike $E^2$, the tiling area is an additive quantity, just as the extensive pressure. More importantly, the tiling area adds new insight by relating mechanical equilibrium to geometry. As an additional practical consideration, stress statistics converge faster when the intensive thermodynamic parameter is conjugate to $\cal A$. Finally, a Boltzmann factor of the form $\exp{(-\gamma {\cal A})}$ supplies the physical intuition, confirmed in the force network ensemble, that local stress statistics in two dimensions should display Gaussian tails. In a system with Boltzmann factor $\exp{(-\alpha {\cal P})}$, an ideal gas-like approximation incorrectly predicts exponential tails. We have demonstrated numerically that $P(p)$ in a canonical system with Boltzmann factor $\exp{(-\alpha {\cal P})}$ and local force balance does indeed approach the correct form with a Gaussian tail for sufficiently large system size. To reproduce the Gaussian tail in a calculation employing this Boltzmann factor, however, would require integrating over local force balance constraints in an ``interacting'' (via Newton's third law) system \cite{bulbul}. 

In analytical calculations, it is often difficult or impossible to integrate out local force balance constraints in a large, potentially disordered system. Approximations to local force balance, e.g.~neglecting spatial correlations or using scalar force balance \cite{coppersmith96}, break the mechanism whereby imposing $\langle {\cal P} \rangle$ also produces the correct $\langle {\cal A} \rangle$. In this situation, it becomes helpful to simultaneously impose both pressure and tiling area constraints in an entropy-maximization calculation, as in Ref.~\cite{tighe08b}. In effect one demands that the system satisfy two global constraints, one of which would have followed ``for free'' due to  local constraints, i.e.~force balance, had they been incorporated exactly. Parameters $\alpha$ and $\gamma$ determined this way can no longer be identified with their values in the thermodynamic limit; instead they should simply be viewed as Lagrange multipliers imposing a constraint. The resulting gain is a dramatic improvement in the accuracy of the predicted statistics, as in Fig.~\ref{fig:naive}, even in an ideal gas-like calculation.

\section*{Acknowledgments}
It is a pleasure to thank Martin van Hecke, Silke Henkes, Jacco Snoeijer, and Zorana Zeravcic for helpful discussions. BPT acknowledges financial support from the Dutch physics foundation FOM.

\appendix
\section{Intensive thermodynamic parameters}
Assuming a Boltzmann factor $\exp{(-\alpha {\cal P} - \gamma {\cal A})}$, we can calculate a relationship in the thermodynamic limit between the thermodynamic parameters $\alpha$ and $\gamma$. We take the system to have $N$ grains and $N_{\rm c}$ contacts. The partition function is 
\begin{eqnarray} \label{app:partition}
Z &=& \int {\rm d}{\bf f} \, e^{-\alpha {\cal P}({\bf f}) - \gamma {\cal A}({\bf f})} \Theta({\bf f})  \nonumber \\
 &=& \int {\rm d}{\cal P}\,{\rm d}{\cal A} \, G({\cal P},{\cal A}) \, e^{-\alpha {\cal P} - \gamma {\cal A}} \,.
\end{eqnarray}
$\Theta({\bf f})$ restricts to balanced noncohesive force networks. $G({\cal P},{\cal A})$ is the density of states with extensive pressure $\cal P$ and tiling area $\cal A$, which can be rewritten $G({\cal P},{\cal A}) = \Omega({\cal P})\Psi({\cal A}|{\cal P})$.  $\Omega({\cal P})$ is the density of states with $\cal P$, which scales as $\Omega({\cal P}) \sim {\cal P}^{N_{\rm c} - 2N - 1}$ in a frictionless disk packing \cite{tighe08b}. $\Psi({\cal A}|{\cal P})$ is the conditional density of states with tiling area $\cal A$ given extensive pressure $\cal P$. The thermodynamic parameters $\alpha$ and $\gamma$ must be chosen to ensure 
\begin{eqnarray} \label{app:params}
\langle {\cal P} \rangle &=& -\frac{\partial\, \ln{Z}}{\partial \alpha} \nonumber \\
\langle {\cal A} \rangle &=& -\frac{\partial\, \ln{Z}}{\partial \gamma} \,.
\end{eqnarray}

The conditional probability $\Psi$ is sharply peaked near $\langle {\cal A}({\cal P}) \rangle$, the average area of a tiling with extensive pressure $\cal P$ \cite{inprogress}. As the system size grows, this function is increasingly well approximated by the form $\langle {\cal A}({\cal P}) \rangle \approx C {\cal P}^2$ \cite{tighe08b,inprogress} for some constant $C$. Therefore,
\begin{equation} \label{app:approx}
\int {\rm d}{\cal A}\, \Psi({\cal A}|{\cal P}) \, e^{ - \gamma {\cal A}} \approx
 e^{-\gamma C{\cal P}^2  } \,
\end{equation}
up to a prefactor that can be absorbed in $\Omega$. Inserting  (\ref{app:approx}) into  (\ref{app:partition}), integrating by parts, and simplifying via  (\ref{app:params}) yields  (\ref{eqn:approx}). Though  (\ref{eqn:approx}) is an approximation, its accuracy increases with growing system size. 

In the thermodynamic limit $N_{\rm c}/N - 2 \rightarrow \frac{1}{2}\Delta z$, where $\Delta z = z - z_{\rm iso}$. Therefore, as we must have either $\alpha \rightarrow 0$ or $\gamma \rightarrow 0$ in the thermodynamic limit,  (\ref{eqn:approx}) predicts the limiting values $\alpha \langle p \rangle = \frac{1}{2}\Delta z$ ($\gamma\rightarrow 0$) and $\gamma \langle a \rangle = \frac{1}{4}\Delta z$ ($\alpha\rightarrow 0$), consistent with the insets of Fig.~\ref{fig:numerics1}. Note that $\alpha$ or $\gamma$ vanish, and hence statistics can no longer be normalized, when $z \downarrow z_{\rm iso}$. This is because the force network ensemble vanishes at isostaticity. 


\end{document}